# Magnetic Kagome Superconductor CeRu$_2$

L. Z. Deng[1], M. Gooch[1], H. X. Liu[2,3], T. Bontke[1], J. Y. You[4], S. Shao[5], J. X. Yin[6], D. Schulze[1], Y. G. Shi[2,3], Y. P. Feng[4,7], G. Chang[5], Q. M. Si[8], C. W. Chu[1,9]

1. Texas Center for Superconductivity and Department of Physics, University of Houston, Houston, Texas 77204, USA

2. Beijing National Laboratory for Condensed Matter Physics and Institute of Physics, Chinese Academy of Sciences, Beijing 100190, China

3. University of Chinese Academy of Sciences, Beijing 100049, China

4. Department of Physics, National University of Singapore, Singapore 117551, Singapore

5. Division of Physics and Applied Physics, School of Physical and Mathematical Sciences, Nanyang Technological University, Singapore 637371, Singapore

6. Laboratory for Topological Quantum Matter and Advanced Spectroscopy (B7), Department of Physics, Princeton University, Princeton, New Jersey 08544, USA

7. Centre for Advanced 2D Materials, National University of Singapore, Singapore 117546, Singapore

8. Department of Physics & Astronomy, Rice Center for Quantum Materials, Rice University, Houston, Texas 77005, USA

9. Lawrence Berkeley National Laboratory, Berkeley, California 94720, USA

Abstract:

Materials with a kagome lattice provide a platform for searching for new electronic phases and investigating the interplay between correlation and topology. Various probes have recently shown that the kagome lattice can host diverse quantum phases with intertwined orders, including charge density wave states, bond density wave states, chiral charge order, and, rarely, superconductivity. However, reports of the coexistence of superconductivity and magnetic order in kagome materials remain elusive. Here we revisit a magnetic superconductor CeRu$_2$ with a kagome network formed by Ru atoms. Our first-principles calculations revealed a kagome flat band near the Fermi surface, indicative of flat-band magnetism. At ambient pressure, CeRu$_2$ exhibits a superconducting transition temperature ($T_c$) up to ~ 6 K and a magnetic order at ~ 40 K. Notably, superconductivity and related behavior can be tuned by adjusting the amount of Ru. We conducted a systematic investigation of the superconductivity and magnetic order in CeRu$_2$ *via* magnetic, resistivity, and structural measurements under pressure up to ~ 168 GPa. An unusual phase diagram that suggests an intriguing interplay between the compound's superconducting order parameters has been constructed. A $T_c$ resurgence was observed above pressure of ~ 28 GPa, accompanied by the sudden appearance of a secondary superconducting transition. Our experiments have identified tantalizing phase transitions driven by high pressure and suggest that the superconductivity and magnetism in CeRu$_2$ are strongly intertwined.

I.  Introduction

The intertwining of superconductivity, magnetism, and non-trivial topology is at the frontier of condensed matter physics. $CeRu_2$ is a magnetic superconductor with a kagome lattice that was discovered over 60 years ago and may have been one of the first among what are now called "kagome superconductors" [1], but its connection to kagome lattice physics has long been overlooked. The pioneering kagome model has predicted many exotic features of magnetic kagome superconductors [2], and several recent experiments have shown hints of time-reversal symmetry breaking in kagome superconductors [3-6]. It is thus timely to revisit the magnetic kagome superconductor $CeRu_2$, in the kagome lattice of which time-reversal symmetry explicitly breaks. In $CeRu_2$, Ru atoms form a kagome lattice perpendicular to the [111] direction (Fig. 1A). Our density-functional theory (DFT) calculations revealed a kagome flat band near the Fermi surface (Fig. 1B-D). Recent angle-resolved photoemission spectroscopy (ARPES) measurements further support our findings since flat bands were detected in this system [7]. Initial studies of $CeRu_2$ can be traced back to B. T. Matthias *et al.*, who in 1958 were searching for ferromagnetic superconductors among solid solutions of $CeRu_2$ and $PrRu_2$ or $GdRu_2$ [1]. It was suggested that for $(Ce,Pr)Ru_2$, there is a region below 1 K where superconductivity and ferromagnetism might overlap. This system has also been found to exhibit reentrant superconductivity based on the observation of a second hysteresis loop between 3.6 T and 5.4 T during a magnetization *vs.* magnetic field measurement at 2.2 K [8]. However, it was later pointed out that such hysteresis is due to a Fulde–Ferrell–Larkin–Ovchinnikov (FFLO) state rather than reentrant superconductivity [9].

To date, only a few materials with a kagome lattice have been found to be superconducting, such as the vanadium-based layered kagome material $AV_3Sb_5$ (A = K, Rb, and Cs) with a superconducting transition temperature ($T_c$) up to ~ 2.5 K [10]; the intermetallic "132" phase $LaRu_3Si_2$ [11, 12], $LaIr_3Ga_2$ [13], *etc.* with a $T_c$ up to ~ 7 K; and $CeRu_2$ with a $T_c$ up to ~ 6 K [1, 8, 14, 15]. The kagome material $Rb_2Pd_3Se_4$ was also recently reported to be superconducting with a $T_c$ up to ~ 3.6 K, but superconductivity only emerges in the compound under pressure (> 20 GPa) [16]. Among the small set of currently known kagome superconductors, $CeRu_2$ sets itself apart as the only one known to exhibit magnetic order. Additionally, high pressure plays a significant role in superconductivity research [17, 18] since it can change the basic structure, tune the carrier concentration, shift the Fermi level, and reshape the Fermi surface topology of a material without altering its chemistry. Therefore, we decided to revisit the $CeRu_2$ system by performing systematic resistive, magnetic, and structural measurements under high pressure, along with theoretical calculations, to explore its superconductivity and magnetism and their possible correlation to the flat bands related to its kagome lattice.

II.  Results

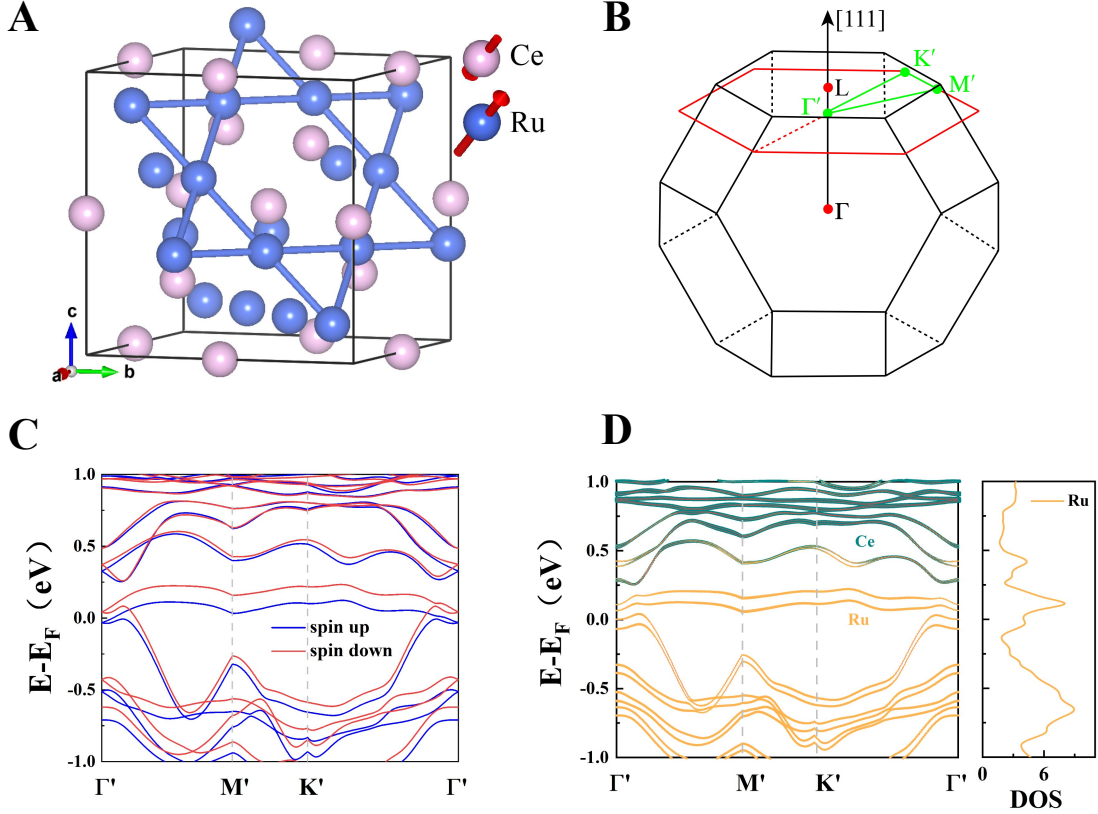

Fig. 1. (A) Crystal structure of CeRu$_2$. Ru atoms form a kagome lattice in the plane perpendicular to the [111] direction. The Ru and Ce atoms have opposite spin magnetic moments, 0.097 and -0.013 μ$_B$, respectively. (B) First Brillouin zone with the specific paths indicated. (C) Band structure without spin-orbit coupling (SOC). (D) Band structure with SOC and corresponding partial density of states (PDOS).

The crystal structure of CeRu$_2$ is presented in Fig. 1A, which shows that the Ru atoms form a kagome lattice in the plane perpendicular to the [111] direction. According to the crystal symmetries, Ru atoms form a kagome lattice in the planes perpendicular to the [11$\bar{1}$], [1$\bar{1}$1], and [$\bar{1}$11] directions. The spin-polarized band structure along the specific paths with Γ'(2/5,2/5,2/5), M'(11/15,7/30,7/30), and K'(11/15,2/5,1/15) (see Fig. 1B) are plotted in Fig. 1C. It should be noted that the spin splitting in the relatively low energy region corresponding to the contribution of the Ru atoms is larger than that in the relatively high energy region corresponding to the contribution of the Ce atoms. This is consistent with the fact that the spin magnetic moment of Ru atoms (0.097 μ$_B$) is larger than that of Ce atoms (-0.013 μ$_B$). We can also see a pair of spin-splitting flat bands near the Fermi level, which are degenerate with large dispersion bands at the Γ' point. These bands are mainly dominated by the Ru atoms and exhibit the band characteristics of the kagome lattice, except that the Weyl point at the K' point is removed due to the complex electron hopping and strong multi-orbital hybridization. The SOC will lift the band degeneracy at the Γ' point, leaving two flat bands that contribute a large density of states (DOS) peak near the Fermi level as shown in Fig. 1D.

Systematic measurements were performed on CeRu$_{2-x}$ single crystals with different amounts of Ru: CeRu$_{1.90}$ and CeRu$_{1.76}$ with T$_c$ ~ 5.1 K and ~ 6.1 K, respectively, at ambient pressure (Fig. 2A). Previous studies have reported different $T_c$s for $CeRu_{2-x}$, but the variation has never been linked to the amount of Ru [1, 8, 9, 15]. The varying amount of Ru may affect the valence of cerium in this compound [19]. We suggest that the main cause for the variation in the $T_c$ of $CeRu_{2-x}$ is the amount of Ru, which may also lead to other changes observed in the physical properties. For example, the $CeRu_{2-x}$ single crystal with a lower amount of Ru has a ~ 22% higher $T_c$ and a ~ 26% higher $H_{C2}$ (Fig. 2B), while that with a higher amount of Ru exhibits a more prominent hysteresis above 2 T (Fig. 2A). As shown in Fig. 2C, the field-dependent resistivity measurement of $CeRu_{1.90}$ at 2 K further reveals that the sample remains superconducting at 2 K below 4.3 T, indicating that the second hysteresis loop observed (Fig. 2A) is not due to reentrant superconductivity. Interestingly, dR/dT analysis of $CeRu_{1.76}$ showed a peak temperature of ~ 38 K at ambient pressure (Fig. 2D), which provides us a way to track the magnetic order in this system. It was previously reported that a magnetic transition was observed at ~ 40 K by muon-spin-relaxation and ac low-field magnetization measurements [20] and very recent muon spin rotation studies [21]. This magnetic order temperature remains constant under magnetic fields up to 7 T (Fig. 2D).

Fig. 3 shows the pressure effect under hydrostatic conditions for magnetization measurements. Notably, the superconductivity volume fraction of $CeRu_{1.90}$ increases with increasing pressure. In addition, the pressure effect on the second hysteresis region is prominent for $CeRu_{1.90}$ and the size and field region of the hysteresis increases with increasing pressure. Measurements were also carried out on $CeRu_{1.76}$ under the same conditions, but no clear trend was detected (Fig. S1). For both $CeRu_{1.90}$ and $CeRu_{1.76}$ single crystals, the $T_c$ remains constant at pressures up to ~ 1.3 GPa.

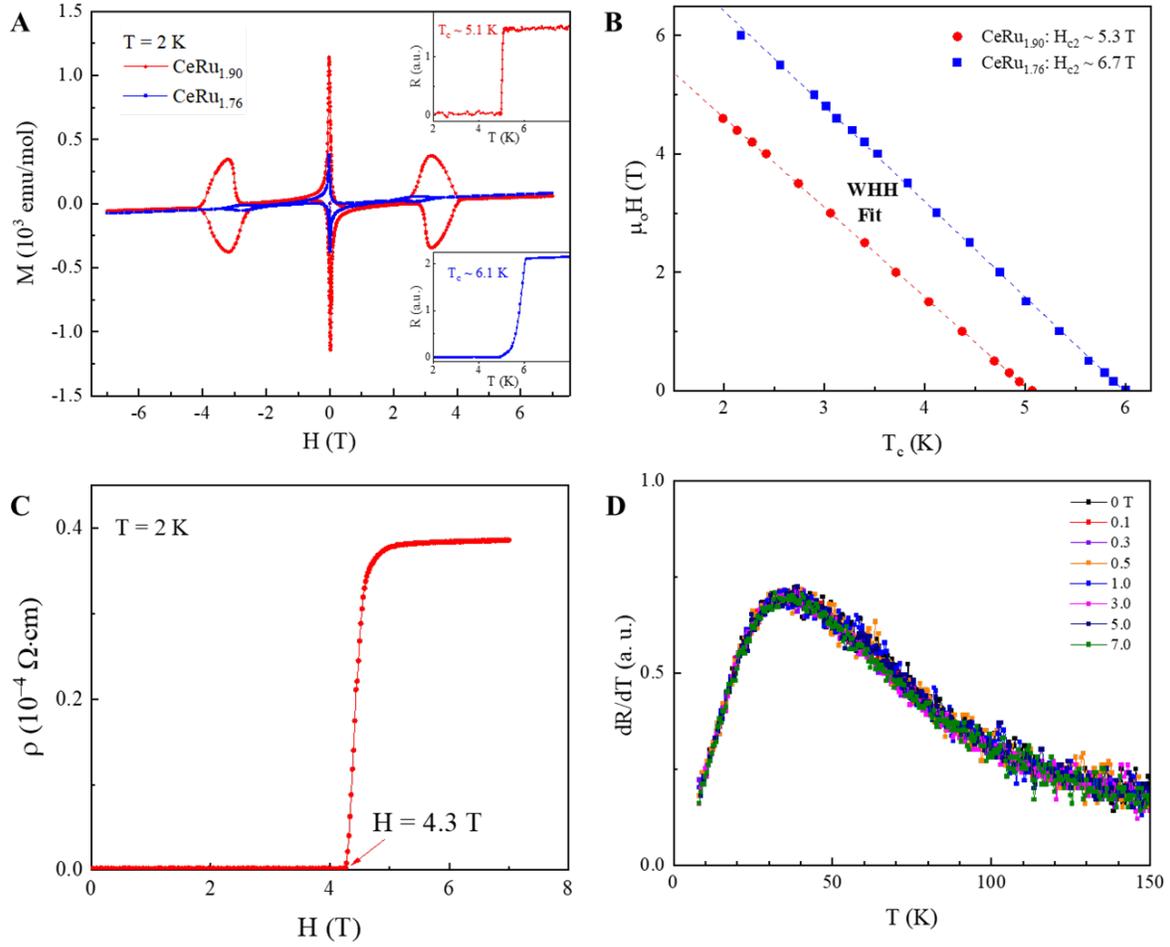

Fig. 2. (A) Magnetization as a function of magnetic field up to 7 T at 2 K for $CeRu_{1.90}$ and $CeRu_{1.76}$. Insets: resistance of $CeRu_{1.90}$ (upper) and $CeRu_{1.76}$ (lower) as a function of temperature. (B) Upper critical field $\mu_0 H_{c2}(T)$ as a function of superconducting critical temperature $T_c^{90\%}$ for $CeRu_{1.90}$ and $CeRu_{1.76}$. The dashed lines represents Werthamer-Helfand-Hohenberg (WHH) fitting [22]. (C) Resistivity as a function of magnetic field up to 7 T at 2 K for $CeRu_{1.90}$. (D) dR/dT under different magnetic fields up to 7 T for $CeRu_{1.76}$.

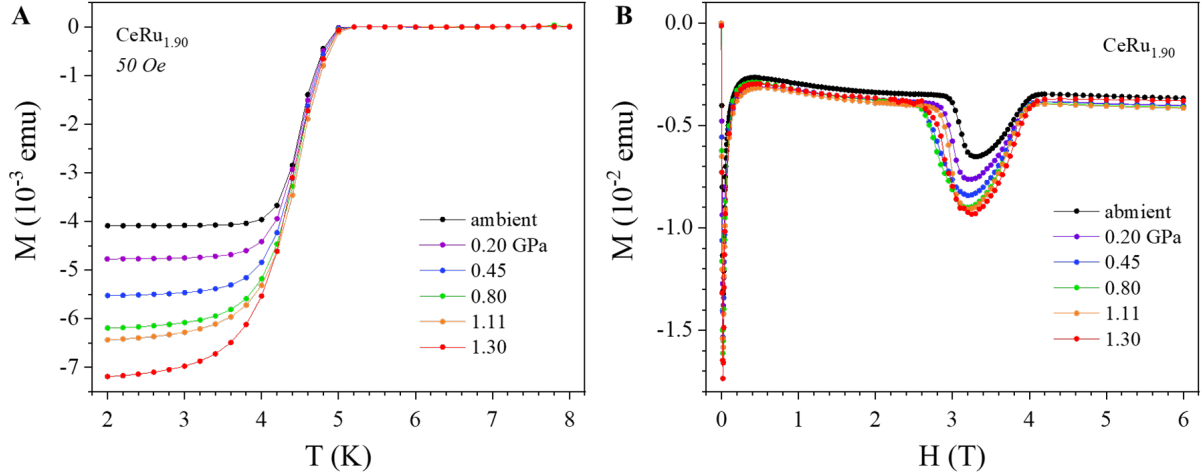

Fig. 3. Magnetic measurements under pressure up to 1.3 GPa for $CeRu_{1.90}$. (A) Magnetic moment as a function of temperature under 50 Oe at different pressures. (B) Magnetic moment as a function of magnetic field at 2 K under different pressures.

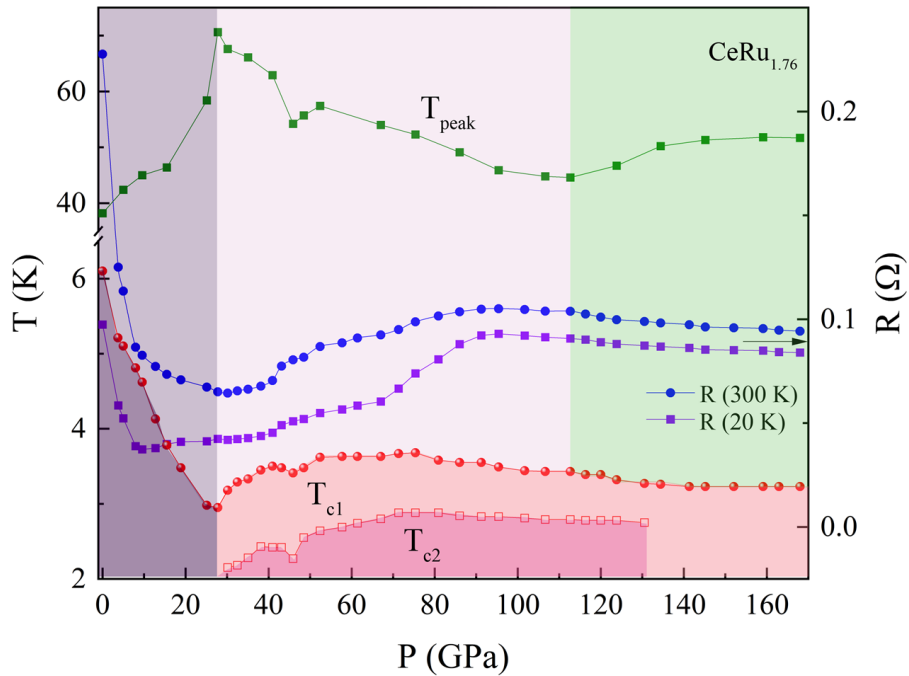

Fig. 4. Phase diagram for $CeRu_{1.76}$. Red circles represent the onset $T_c$ and red squares represent a secondary $T_c$ that emerges at ~30 GPa. Blue circles and violet squares represent resistance measured at room temperature and at 20 K, respectively. Green squares show the peak position of $dR/dT$.

Resistivity measurements were performed under pressures up to 168 GPa for single-crystal $CeRu_{1.76}$. The phase diagram is shown in Fig. 4, and detailed resistance vs. pressure data is shown in Fig. 5. Both room-temperature (RT) resistance and $T_{c1}$ decrease with increasing pressure up to ~ 28 GPa. Above 28 GPa, an increase in the $T_{c1}$ is detected, indicating a possible phase transition. Simultaneously, a secondary $T_{c2}$ is observed as a step at a lower temperature. With further increasing pressure, a decrease in $T_{c1}$ begins above

~ 75 GPa, its rate of reduction decreases above 113 GPa, and it remains nearly constant from ~ 141 GPa to ~ 168 GPa, the highest pressure reached in this experiment. It is interesting to note that the $T_{peak}$ from dR/dT evolves with pressure and correlates well with the observed behavior of the $T_c$ under pressure. Overall, the magnetic order seems to compete with the superconductivity since they have opposite pressure effects. However, between 75 GPa and 113 GPa, both the superconducting $T_c$ and the magnetic order $T_{peak}$ decrease with increasing pressure. Resistance measurements on single-crystal $CeRu_{1.90}$ were also carried out under pressure up to ~ 50 GPa, as shown in Fig. 6. Similar to the case of $CeRu_{1.76}$, a $T_c$ resurgence is observed at pressures above ~ 28 GPa. The $T_c$-P slope changes at ~ 7 GPa, which correlates to the transition temperatures of R(20 K) *vs.* P and R(300 K) *vs.* P, indicating another possible phase transition. It was previously reported that a pressure-induced phase transition occurs above 5.5 GPa [23]. Overall, the $T_{peak}$, indicating the magnetic order temperature, varies with pressure in an opposite way from that of the superconducting $T_c$: from ambient pressure to ~ 27.5 GPa, $T_{peak}$ increases with increasing pressure while $T_c$ simultaneously decreases, and above 28 GPa and up to ~ 48 GPa, $T_{peak}$ decreases and $T_c$ increases with increasing pressure. It is worth mentioning that a drop is observed in the $T_{peak}$ *vs.* P between ~ 6 GPa and 10 GPa, possibly due to the phase transition [23]. Our pressure measurement results suggest that superconductivity and magnetism are strongly intertwined in $CeRu_2$.

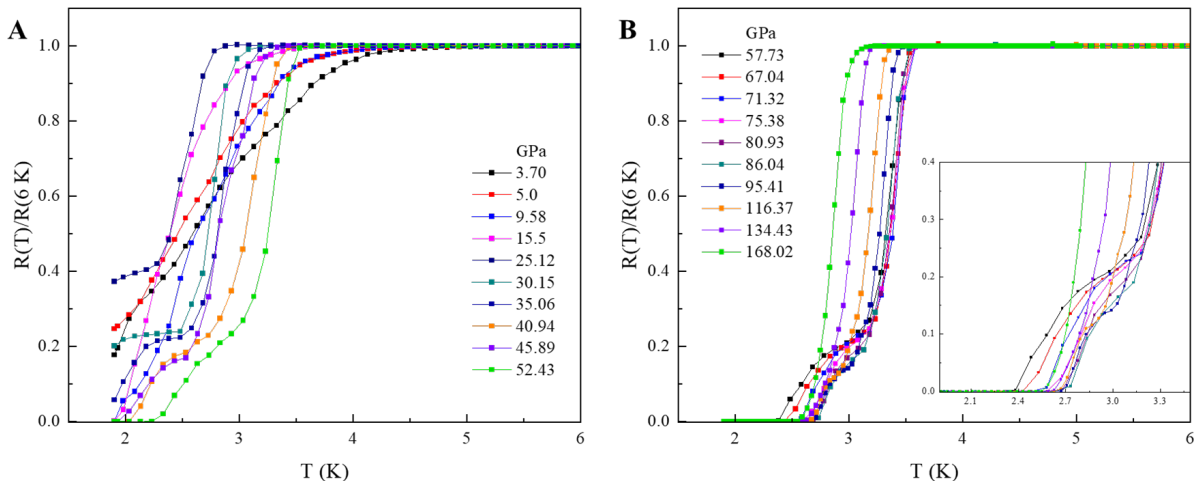

Fig. 5. Temperature-dependent resistance of $CeRu_{1.76}$ at various pressures in the ranges of (A) 3.70 GPa to 52.43 GPa and (B) 57.73 GPa and 168.02 GPa. Inset to (B): magnified view of the low-resistance region at higher pressures.

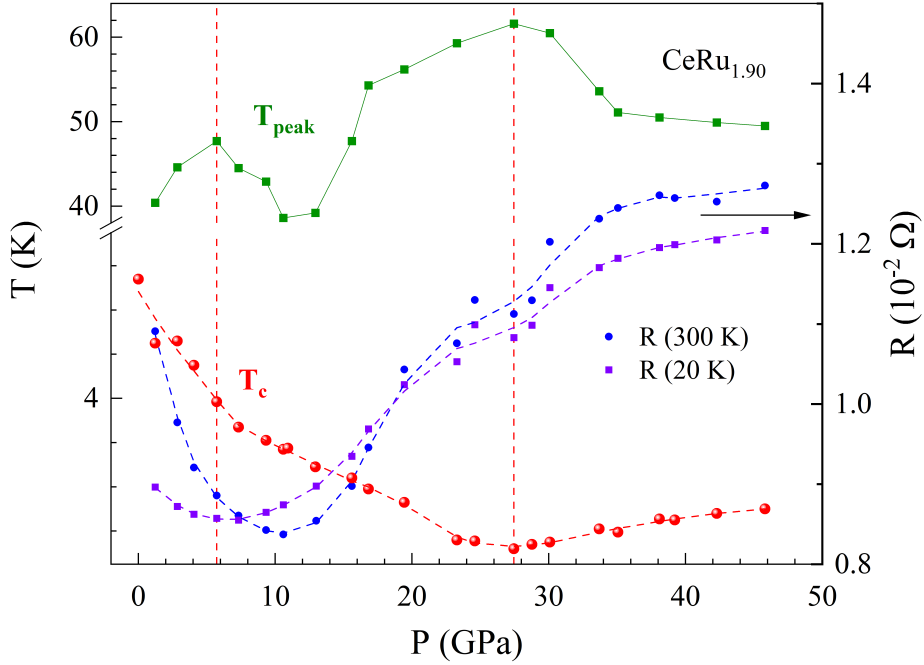

Fig. 6. Phase diagram for CeRu$_{1.90}$. Red circles denote the onset T$_c$. Blue and violet solid symbols represent resistance measured at room temperature and 20 K, respectively. Green squares show the peak position of dR/dT. R (20K) *vs.* P was shifted up vertically by 0.006 Ω. Dashed lines are guides for the eye.

III.     Conclusion

The magnetic superconductor CeRu$_2$ hosts a kagome network formed by Ru atoms. By examining CeRu$_{2-x}$ single crystals with different amounts of Ru, we discovered that the varying amount of Ru may be the main cause for the different T$_c$s have been reported for this system and demonstrated that it has significant impacts on the second hysteresis loop observed in this system. Through systematic high-pressure investigations on this system under pressure up to ~ 168 GPa, we discovered a T$_c$ resurgence and a secondary superconducting transition above 28 GPa, a few possible phase transitions induced by high pressure that may be either structural or electronic, and a magnetic order that can be tuned by pressure. We further demonstrated that superconductivity and magnetic order are strongly intertwined in this system.

**Methods**

**Sample Preparation**. Polycrystalline CeRu$_2$ was synthesized by arc melting stoichiometric amounts of Ce and Ru. The single crystals of CeRu$_{2-x}$ were grown using the Czochralski pulling method in a tetra-arc furnace. First, a polycrystalline ingot 10 g in mass was prepared in an arc furnace as a precursor. The ingot was then placed on the water-cooled copper hearth of the tetra-arc furnace under an Ar atmosphere with a Ti ingot for adsorbing oxygen. A current of 16 A was applied to the ingot to melt it and the hearth rotated at a speed of 0.3 revolutions per minute. A columnar single crystal was successfully grown after several hours. Chemical composition was determined by energy-dispersive spectroscopy (EDS) using a JEOL JSM-6330F scanning electron microscope with an energy-dispersive X-ray spectrometer.

**Electrical Transport Measurements under Pressure.** For resistivity measurements conducted in this investigation, the pressure was applied to the samples using a symmetric-type diamond anvil cell. We used anvils with a 150 μm central culet beveled to 300 μm at 8°. The rhenium gasket was insulated with Stycast 2850FT. The sample's chamber diameter was ~ 80 μm, where cubic boron nitride (cBN) is used as the pressure-transmitting medium (PTM). Samples were cleaved and cut into thin squares with a diagonal of ~ 70 μm and thickness of ~ 10 μm. The pressure was determined using the ruby fluorescence scale [24] or the diamond Raman scale at room temperature [25]. The samples' contacts were arranged in a Van der Pauw configuration and data were collected using a Quantum Design Physical Property Measurement System (PPMS) with temperatures down to 1.9 K and magnetic fields up to 7 T.

**Magnetic Measurements under Pressure.** A piston-cylinder-type high-pressure cell, compatible with the Quantum Design Magnetic Property Measurement System (MPMS), was used when performing low-pressure measurements up to 1.3 GPa, where the pressure medium was Daphne-7373 oil.

**Theoretical Calculations.** Our first-principles calculations were based on density-functional theory (DFT) as implemented in the Vienna Ab initio Simulation Package (VASP) [26, 27], using the projector augmented-wave method [28]. The generalized gradient approximation with the Perdew-Burke-Ernzerhof solid [29] realization was adopted for the exchange-correlation functional. The plane-wave cutoff energy was set to 500 eV. A Monkhorst-Pack *k*-point mesh [30] with a size of 15×15×15 was used for the Brillouin zone sampling. The crystal structure was optimized until the forces on the ions were less than 0.001 eV/Å. To account for the correlation effects for transition-metal elements, the DFT+*U* method with $U$=7 eV and 1.65 eV for Ce and Ru atoms, respectively, was used for calculating the band structures, which showed magnetism consistent with the experiment.

**Acknowledgments.** L.Z.D., M.G, T.B., D.S., and C.W.C are supported by US Air Force Office of Scientific Research Grants FA9550-15-1-0236 and FA9550-20-1-0068, the T. L. L. Temple Foundation, the John J. and Rebecca Moores Endowment, and the State of Texas through the Texas Center for Superconductivity at the University of Houston. S. S. and G. C. are supported by the National Research Foundation, Singapore, under its Fellowship Award (NRF-NRFF13-2021-0010) and the Nanyang Assistant Professorship grant from Nanyang Technological University. J.Y.Y. and Y.P.F. are supported by the Ministry of Education, Singapore, under its MOE AcRF Tier 3 Award MOE2018-T3-1-002.